\documentclass[review,sort&compress]{elsarticle}

\usepackage{lineno,hyperref}
\usepackage{amsmath,amssymb,amsfonts}
\usepackage{algorithmic}
\usepackage{graphicx}
\usepackage{textcomp}
\usepackage{xcolor}
\usepackage{algorithm}
\usepackage[section]{placeins}
\usepackage{setspace}
\usepackage{booktabs}

\newcommand{\RNum}[1]{\uppercase\expandafter{\romannumeral #1\relax}}

\journal{Journal of \LaTeX\ Templates}

\begin{document}

\begin{frontmatter}

\title{An Iterative Graph Spectral Subtraction Method for Speech Enhancement\tnoteref{mytitlenote}}
\tnotetext[mytitlenote]{This work has been supported by National Natural Science Foundations of China\\(Grant Nos.: 61671252 and 61271335)}

\author[mymainaddress]{Xue Yan\corref{mycorrespondingauthor}}
\cortext[mycorrespondingauthor]{Corresponding author}
\ead{1018010432@njupt.edu.cn}

\author[mymainaddress,mysecondaryaddress]{Zhen Yang}
\ead{yangz@njupt.edu.cn}
\author[mymainaddress]{Tingting Wang} 
\ead{2018010215@njupt.edu.cn}
\author[mymainaddress]{Haiyan Guo} 
\ead{guohy@njupt.edu.cn}

\address[mymainaddress]{Department of Communication and Information Engineering, Nanjing University of Posts and Telecommunications}
\address[mysecondaryaddress]{National Engineering Research Center of Communication and Sensor Network Technology}

\begin{abstract}
In this paper, we investigate the application of graph signal processing (GSP) theory in speech enhancement. We first propose a set of shift operators to construct graph speech signals, and then analyze their spectrum in the graph Fourier domain. By leveraging the differences between the spectrum of graph speech and graph noise signals, we further propose the graph spectral subtraction (GSS) method to suppress the noise interference in noisy speech. Moreover, based on GSS, we propose the iterative graph spectral subtraction (IGSS) method to further improve the speech enhancement performance. Our experimental results show that the proposed operators are suitable for graph speech signals, and the proposed methods outperform the traditional basic spectral subtraction (BSS) method and iterative basic spectral subtraction (IBSS) method in terms of both signal-to-noise ratios (SNR) and mean Perceptual Evaluation of Speech Quality (PESQ).
\end{abstract}

\begin{keyword}
graph signal processing\sep  speech enhancement\sep  graph Fourier transform\sep  graph spectral subtraction
\end{keyword}

\end{frontmatter}

\section{Introduction}
In recent years, GSP has attracted extensive attention as a new signal processing method via developing and utilizing the correlations or some other information among signals\cite{b1,b2,b3}. It extends classical discrete signal processing (DSP) to signals indexed by graphs\cite{b4}. GSP has been widely used in wireless sensor networks (WSN) \cite{b5,b6,b7}, image processing,\cite{b8,b9,b10}, acoustic scene analysis\cite{b33}, machine learning\cite{b11}, etc. It provides a new way to process data, where the correlations among data are considered in the form of edges and weights. GSP shows good performance especially in handling large, irregular data.

The GSP theory can be traced back to the Algebra Signal Processing (ASP) theory\cite{b12,b13}, which provides a method to visualize signal models. The key insight of ASP is to identify the shift operator, which can be seen as the weighted matrix $\cal{W}$ of the visualized graph signal. By defining $\cal{W}$, GSP takes into account both the influence of the current vertex and the adjacent vertices to describe the intrinsic relations among vertices. This is different from the traditional DSP methods where only the current vertex is considered.  

There are two typical representations of $\cal{W}$ for graph signals. One is the graph Laplacian matrix  $\cal{L}$ used in \cite{b14,b15,b16}, and the other is the graph adjacency matrix $\cal{A}$ used in  \cite{b32,b17,b18}. The generally used combinatorial graph Laplacian matrix $\cal{L}$ is defined as ${\cal{L = D - A}}$, where the degree matrix $\cal{D}$ is a diagonal matrix derived from $\cal{A}$ (whose elements are denoted as ${a_{ij} }$) with the diagonal element $d_{ii}= \sum\nolimits_{i=1}^N {a_{ij} }$. Note that methods for $\cal{L}$-based graph signal analysis can only be used for undirected graphs, while $\cal{A}$ does not have such limitation \cite{b3}. Considering that speech signals are directed time series, in this paper, we mainly focus on the directed graph and view $\cal{A}$ as our $\cal{W}$ to investigate its construction.
 
Once $\cal{A}$ is given or estimated, operations such as graph Fourier transform (GFT) and filtering can be extended. As presented in \cite{b3}, $\cal{A}$ plays a dual role which represents both the shift operator $z^{-1}$ in DSP and the adjacency matrix of a cycle graph. In \cite{b19}, the authors used  the sparsity of graph spectrum for image compression, and defined the graph Fourier basis as the generalized eigenvectors of $\cal{A}$ to perform GFT and obtain the corresponding graph spectrum. In \cite{b20}, the authors proposed a data classification method by interpreting the classifier system as a graph filter and discussed the properties of graph filters, including linearity, shift-invariance and invertibility. In \cite{b21}, the authors defined a set of energy-preserving shift operators, and proved that $\cal{A}$ is a linear shift invariant (LSI) graph filter with respect to the defined operator. 

Although GSP has made outstanding achievements in many fields, it is mainly used to deal with large-scale or irregular data. To our best knowledge, there have been few specific GSP works on low dimensional signals such as speech. This is mainly because that the graph, which describes their internal relationships among data, cannot be obtained directly from an existing topology of the low dimensional signal. On the other hand, GSP has great potential in speech processing. Firstly, GSP directly describes the relationships among vertices with edges and weights when constructing the graph. And the direction of edges further describes the relationships more deeply. As a time series, there obviously exist correlations among the adjacent vertices of speech signal, which may appropriate for such an representation by graph. Secondly, the graph frequency domain is variable, the adjacency matrix constructed by different edges and weights defines different graph Fourier transforms. Different domains can separate the useful speech from the useless noise in different degrees, thus one can map speech signals to different graph frequency domains according to different needs of processing.

Speech enhancement technology, which aims to improve the speech quality, is considered as a necessary preprocessing step for many speech applications since the input speech is generally not so pure\cite{b22}. The traditional DSP based speech enhancement methods mainly include spectral subtraction methods \cite{b23,b24}, methods based on statistical features \cite{b25,b26,b27}, subspace based methods \cite{b28} and machine learning based methods \cite{b29}, etc. 

Motivated by the advantages of GSP and the need of speech enhancement, we investigate the application of GSP theory in speech enhancement and propose graph spectral subtraction (GSS) method and the iterative graph spectral subtraction (IGSS) method to suppress the noise interference in noisy speech based on the assumption that speech and noise are statistically independent. The main contributions of this paper are summarized as follows.
\begin{enumerate}[1)]
\item We proposed a set of combined $k$-shift operators to map speech signals to the graph domain.
\item We propose the GSS and IGSS methods for speech enhancement based on the characteristics of speech signals in the corresponding graph frequency domain.
\item Our experimental results show the proposed combined shift operators are suitable for graph speech signals and the proposed IGSS method leads to higher SNR and PESQ than the traditional spectral  subtraction methods.

\end{enumerate}

The outline of this paper is as follows. In section \ref{S2}, the construction of speech signals on graphs by leveraging the proposed shift operators is discussed. After which, in section \ref{S3}, the GSS and IGSS methods are presented. Our experimental results are provided in section \ref{S4}, while section \ref{S5} concludes this paper.

\section{Speech Signals on Graphs}\label{S2}

\subsection{Speech signal in graph domain}\label{2A}
In this subsection, we introduce the basic concept of graph signal and construct the graph speech signal. Graph in GSP describes the relationships among vertices, which is generally defined as $\cal {G = (V, E, W)}$\cite{b1}. Elements $\cal {V, E}$ and $\cal W$ represent the vertex set, the edge matrix and the weighted matrix, respectively. To apply the GSP processing approach to speech signal, we first need to map a speech signal $s$ in time domain to the graph speech signal $\bf{s}_\mathcal{G}$ in graph domain. In this paper, the speech signal is processed in frames. Considering a speech frame $s=[s_{n_0},s_{n_1}, ..., s_{n_{N-1}}]^{\rm T}$ with $N$ points, we can represent it as a numerical-valued signal $\bf{s}_\mathcal{G}$ indexed by a graph $\mathcal{G}$. 
\begin{equation}\label{eq1}
p:\quad s \to {{\bf{s}}_\mathcal{G}}\in \mathbb{R}^N \quad \text{\emph{indexed by}}\quad \mathcal{G=(V,E,W)}.
\end{equation}
The relationships among the vertices are denoted as a graph $\mathcal{G=(V,E,W)}$, and  $\mathcal{V}={[v_0,v_1,...,v_{N-1}]}^{\rm{T}}$ is the set of all vertices of graph. Each vertex $v_k$ corresponds to a sampling point $n_k$ in time domain. ${\bf{s}_\mathcal{G}}=[s_{\mathcal{G}_0},s_{\mathcal{G}_1}, ..., s_{\mathcal{G}_{N-1}}]^{\rm{T}}$ is the one-to-one map of $s$, and each $s_{\mathcal{G}_k}$ represents the intensity of the corresponding $v_k$.  $\mathcal{E}\in {{\mathbb{R}}^{N\times N}}$ is the set of edges with ${e_{ij}} \in \left\{ {0,1} \right\}$, and $\mathcal{W}={{\left\{ {{\omega }_{ij}} \right\}}_{\left( i,j \right)\in \mathcal{E}}}\in {{\mathbb{R}}^{N\times N}}$ is the weighted matrix. More specifically, ${{e}_{ij}}=1$ denotes that there exists an edge connecting vertices ${{v}_{i}}$ and ${{v}_{j}}$, otherwise ${{e}_{ij}}=0$. ${{\omega }_{ij}}>0$ denotes the weight of edge from ${{v}_{j}}$ to ${{v}_{i}}$. 

Note that when it comes to an undirected graph, $\mathcal{E}$ and $\mathcal{W}$ are both symmetric matrices. In this paper, we mainly focus on the directed graphs and adopt $\mathcal A$ as our $\mathcal W$. Moreover, we mainly concentrate on the existence of the relationship among vertices rather than the intensity of that. That is to say, $\mathcal A$ is a 0-1 matrix, and $a_{ij}=1$ means there exists a connection from $v_{j}$ and $v_{i}$, otherwise, $a_{ij}=0$. A more detailed description for ${\cal{A}}$ can be developed in future work using correlation coefficients and other methods. 
\subsection{The combined graph shift operators}\label{2B}
In this subsection, we propose a set of combined $k$-shift operators to construct our graph speech signal $\bf{s}_\mathcal{G}$. \emph{Definition 2} in \cite{b12} provides a method to obtain a visualized graph via shift operators. An appropriate graph shift operator for GS, which is similar to shift operator in DSP, is of great significance in GSP, since almost all operations such as transformation, filtering, and prediction are directly related to it. 

Since the speech signal, which is a time sequence with obvious temporal correlations, has a directed weight that represents the relationships among speech samples. Furthermore, inspired by our previous studies on compression sensing, where the row echelon measurement matrix was proposed and proved effective in both suppressing the noise and compressing simultaneously\cite{b30}, we define a novel combined $k$-shift operator $\Psi_k $ and leverage it as the adjacency matrix of the graph of graph speech signal.

Firstly, a $k$-shift operator can be denoted as a 0-1 matrix ${{\Phi}_k}\in {{\mathbb R}^{N \times N}} (k=0, 1, ...)$ with the element $\phi_{ij} $ satisfies
\begin{equation} \label{eq3}
\phi_{ij}  = \left\{ \begin{array}{l}
 1,\quad if\quad (j-i) \quad mod\quad N  = k\\
 0,\quad otherwise \\ 
 \end{array} \right..
\end{equation}
For $k=0$, $\Phi_k$ is a unit matrix which means no shift is down on the signal. For $k>0$, the output ${\bf{s}}_{{\cal G}_{out}}$ after the unit $k$-shift operation can be expressed as
\begin{align} 
{\bf{s}}_{{\cal G}_{out}}&={\Phi}_k  \cdot {\bf{s}}_{{\cal G}_{in}}\\\notag
&=\left[ s_{{\cal G}_{k}},s_{{\cal G}_{k+1}} ,...,s_{{\cal G}_{N-1}} ,s_{{\cal G}_{0}} ,...,s_{{\cal G}_{k-1}} \right]^{\rm T} ,k = 1,2, ... .
\end{align}

\begin{figure}[t]
\centerline{\includegraphics[width=0.6\linewidth]{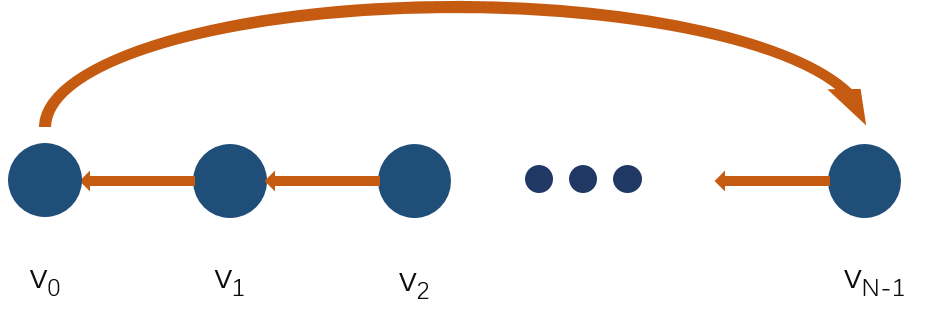}}
\caption{Graph representation for $\mathcal{G}_{\Phi_{1}}=(\mathcal{V},\Phi_{1},\Phi_{1})$.}
\label{Phi1}
\end{figure}
By applying the $k$-shift operator $\Phi_k$ and viewing it as an adjacency matrix, we shift each edge and then obtain connections between corresponding vertices. Thus, the edge sets $\mathcal E$ is then determined. Note that $\cal E$ is equal to $\cal A$ when  $\cal A$ is an 0-1 matrix. So far, we can obtain a graph $\mathcal{G}_{\Phi_k}=(\mathcal{V},\Phi_k,\Phi_k)$ for ${\bf s}_{\mathcal G_{\Phi_k}}$. The visualized graph of $\mathcal{G}_{\Phi_{1}}=(\mathcal{V},\Phi_{1},\Phi_{1})$ is shown in Fig. \ref{Phi1}.

\begin{figure}[t]
\centerline{\includegraphics[width=0.65\linewidth]{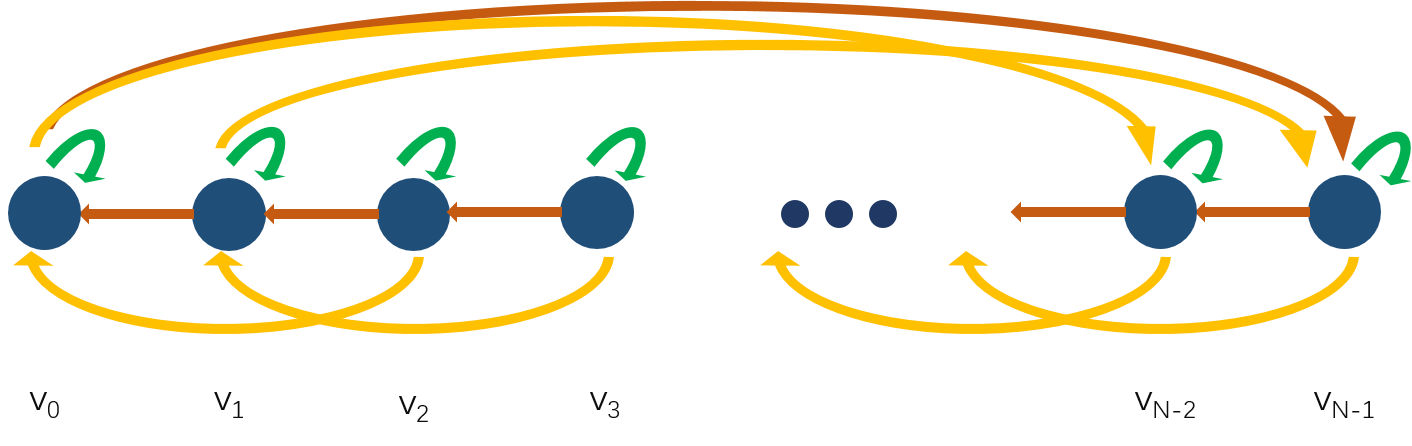}}
\caption{Graph representation for ${\mathcal G}_3=(\mathcal{V},\Psi_{3},\Psi_{3})$ .}
\label{fig2}
\end{figure}

Now we define our combined $k$-shift operator $\Psi_k$ as
\begin{equation} \label{eq6}
\Psi_k = \sum\limits_{i=0}^{k - 1} {\Phi_i } ,k=1, 2, ... ,
\end{equation}
where the element $\psi_{ij} $ satisfies
\begin{equation} \label{eq7}
\psi_{ij}  = \left\{ \begin{array}{l}
 1,\begin{array}{l}if\quad (j-i)\quad mod\quad N = 0,...,k-1 \end{array}\\
 0,\begin{array}{l}otherwise \end{array}
 \end{array} \right..
\end{equation}
Apparently, for $k=1$, there is an $\Psi_{1}=\Phi_0$. The output ${\bf{s}}_{{\cal G}_{out}}$ after the operation of $\Psi_k$ on  ${\bf{s}}_{{\cal G}_{in}}$ can be expressed as ${\bf{s}}_{{\cal G}_{out}}=\Psi_k  \cdot {\bf{s}}_{{\cal G}_{in}}$. More directly, our combined $k$-shift operation consists of two steps. Which are first shifting the edge separately by 0-, 1-, ... , $k-1$-step respectively, and then performing the linear superimposition of each shift operation. We can then get a new $\mathcal{G}_{\Psi_{k}}=(\mathcal{V},\Psi_{k},\Psi_{k})$ when viewing $\Psi_k$ as an adjacency matrix. For simplicity, we use $\mathcal{G}_{k}$ to represent $\mathcal{G}_{\Psi_{k}}$ in the following paper. The graph representation of ${\mathcal G}_3=({\mathcal V},\Psi_{3},\Psi_{3})$ is shown in Fig. \ref{fig2} as an example.

\subsection{Spectrum in graph frequency domain}\label{2C}
Graph signals are generally transformed into the graph frequency domain in order to further analyze their properties. The basis of the graph Fourier transform is derived from the adjacency matrix. By performing eigen-decomposition on $\mathcal A$ , we have

\begin{equation}\label{eq9}
\mathcal A=\varsigma \Lambda {\varsigma} ^{ - 1} ,
\end{equation}
where $\Lambda= diag\left[ {\lambda_0} ,{\lambda_0} ,...,{\lambda_{N-1}}  \right]$ is a diagonal matrix with the diagonal elements being the distinct eigenvalues of $\mathcal A$, and $\lambda _{k}$ represents the graph frequencies. $\varsigma  = \left[ {\nu _0 ,\nu _1 ,...,\nu _{N - 1} } \right]$ is a matrix of $N$ eigenvectors of $\mathcal A$, and the column $\nu _k$ is the spectral component corresponding to $\lambda _{k}$. Since $\mathcal A$ adopted in this paper is a row echelon matrix with a full row rank, which would lead to $N$ linearly independent eigenvectors, we can have that $\varsigma$ is invertible. Thus, the graph Fourier matrix is defined as $\mathcal F$  (following \cite{b19})
\begin{align}\label{eq10}
 \mathcal F &= \varsigma^{-1} =\left[ {\nu _0 ,\nu _1 ,...,\nu _{N - 1} } \right]^{-1} \notag  \\ 
&=\left[ {f_0 ,f_1 ,...,f_{N - 1} } \right]  .
\end{align}
The graph Fourier transform (GFT) of the graph signal $\bf{s}_\mathcal{G}$ can be further defined as ${\bf{\hat s}}_\mathcal{G}$ (following \cite{b19})

\begin{align}\label{eq11}
{\bf{\hat s}}_\mathcal{G} &= \mathcal F \cdot{\bf{ \bf{s}_\mathcal{G}}}= \left[ {f_0 \bf{s}_\mathcal{G}},f_1 {\bf{s}}_\mathcal{G},...,f_{N - 1} \bf{s}_\mathcal{G} \right]^{\rm T}  \notag \\
&=\left[ {\hat s}_{f_0} ,{\hat s}_{f_1},...,{\hat s}_{f_{N-1}} \right]^{\rm T} ,
\end{align}
where ${\hat s}_{f_k}$ denotes the graph frequency coefficient of ${\bf{\hat s}}_\mathcal{G}$ corresponding to each graph frequency $\lambda _k$. The inverse graph Fourier transform (IGFT) (following \cite{b19}) is defined as
\begin{align}\label{eq12}
{\bf{s}}_\mathcal{G} &= {\mathcal F}^{-1} \cdot {\bf{\hat s}}_\mathcal{G}=\varsigma \cdot [ {\hat s}_{f_0} ,{\hat s}_{f_1},...,{\hat s}_{f_{N-1}} ]^{\rm T}\notag \\
&=[s_{\mathcal{G}_0},s_{\mathcal{G}_1}, ..., s_{\mathcal{G}_{N-1}}]^{\rm{T}} .
\end{align}
By utilizing the IGFT, speech signal in graph frequency domain can be transformed into graph domain.

\section{Graph Based Spectral Subtraction Methods}\label{S3}
In this section, we propose the GSS and IGSS methods by utilizing the different distributions between graph speech and graph noise signals in the graph frequency domain.

Considering a noisy speech signal ${y=s+n}$, where $n$ is the additive noise which is statistically independent of the speech, for each speech frame $y^{(m)}$, we construct the noisy graph speech frame ${\bf{y}}^{(m)}$  with $\Psi_k$ via method in (\ref{eq1}). 
\begin{equation}\label {eq13}
p:\quad y^{(m)} \to {\bf{y}}_{{\mathcal G}_k}^{(m)} = {\bf{s}}_{{\mathcal G}_k}^{(m)} + {\bf{n}}_{{\mathcal G}_k}^{(m)}\quad \text{\emph{indexed by}}\quad \mathcal{G}_{k}=(\mathcal{V},\Psi_k,\Psi_k),
\end{equation}
where the additive graph noise frame ${\bf{n}}_{{\mathcal G}_k}^{(m)}$ and the pure graph speech frame ${\bf{s}}_{{\mathcal G}_k}^{(m)}$ are indexed by the same graph ${\mathcal G}_k$. Our purpose is to eliminate the interference of  ${\bf{n}}_{{\mathcal G}_k}^{(m)}$ so as to obtain a pure  ${\bf{s}}_{{\mathcal G}_k}^{(m)}$ from noisy graph speech frame ${\bf{y}}_{{\mathcal G}_k}^{(m)}$. Considering the fact that only noise exists in the non-speech activity area. Moreover, the white noise is relatively stable, which means the statistical characteristics of noise during the speech activity areas are basically the same as that of the non-speech activity area. Therefore, we obtain the estimation of graph noise from the non-speech activity area and then leverage it for removing the component of graph noise from the noisy graph speech. 

Now we present our graph spectral subtraction method (GSS) for speech enhancement. On the consideration of the properties of graph speech and graph noise, our speech enhancement algorithm is performed in the graph frequency domain corresponding to $\Psi_k$. By applying the eigenvalue decomposition of $\Psi_k$ via (\ref{eq9}), we obtain the graph Fourier transform basis ${\mathcal F}_{\Psi_k} = {\varsigma _{\Psi_k}}^{-1}$, then the graph spectral coefficients ${\bf{\hat y}}^{(m)}_{{\mathcal G}_k}$ of ${\bf{y}}_{{\mathcal G}_k}^{(m)}$ can be expressed as
\begin{equation}\label {eq14}
{\bf{\hat y}}^{(m)}_{{\mathcal G}_k} = {\varsigma _{\Psi_k}}^{-1} {\bf{y}}^{(m)}_{{\mathcal G}_k}={\bf{\hat s}}^{(m)}_{{\mathcal G}_k} + {\bf{\hat n}}^{(m)}_{{\mathcal G}_k}.
\end{equation}
\begin{algorithm}[tb]
\setstretch{1.2}
\caption{Graph Spectral Subtraction algorithm (GSS)}
\label{algGSS}
\begin{algorithmic}[1]
\REQUIRE noisy speech $y$; frame length $N$; step $k$ of the combined shift operator $\Psi_k$
\ENSURE the enhanced speech $s_{est}$
\STATE Divide $y$ into $M$ frames with frame length of $N$ and frame shift of $N/2$
\STATE Construct ${\mathcal G}_{k}=({\mathcal V},\Psi_k,\Psi_k)$
\FOR{each frame $y^{(m)}$}
\STATE Construct the ${\bf y}^{(m)}_{{\cal G}_k}\text{{ indexed by }}{\mathcal G}_k$
\STATE Perform GFT on ${\bf y}^{(m)}_{{\cal G}_k}$ to obtain ${\bf{\hat y}}^{(m)}_{{\cal G}_k}$
\STATE  Save the phase $< {\bf{\hat y}}^{(m)}_{{\cal G}_k}> $, and extract the amplitude $| {\bf{\hat y}}^{(m)}_{{\cal G}_k}|^{'} $ of half vertices

\STATE Obtain the estimation of the noise $|{\bf{\hat n}}_{{\cal G}_k}|_{est}^{'}$ and subtract it to get $| {\bf{\hat s}}^{(m)}_{{\cal G}_k}|_{est}^{'}  = | {\bf{\hat y}}^{(m)}_{{\cal G}_k}|^{'} - |{\bf{\hat n}}_{{\cal G}_k}|_{est}^{'}$ 
\STATE Expand $| {\bf{\hat s}}^{(m)}_{{\cal G}_k}|_{est}^{'} $ to  $| {\bf{\hat s}}^{(m)}_{{\cal G}_k}|_{est} $
\STATE Recover $ {\bf{\hat s}}^{(m)}_{{\cal G}_k est}$ with the help of $< {\bf{\hat y}}^{(m)}_{{\cal G}_k}> $
\STATE Perform IGFT on  $ {\bf{\hat s}}^{(m)}_{{\cal G}_k est}$  to get  $ {\bf s}^{(m)}_{{\cal G}_k est}$ 
\STATE Obtain $ {s}^{(m)}_{est}$ by removing ${\mathcal G}_{k}$ of  $ {\bf s}^{(m)}_{{\cal G}_k est}$
\ENDFOR
\STATE Restore $M$ frames to $s_{est}$
\RETURN ${s_{est}}$
\end{algorithmic}
\end{algorithm}
Considering the fact that human ear is insensitive to the phases of speech signals, we only deal with the amplitude spectrum $|{\bf{\hat y}}^{(m)}_{{\mathcal G}_k}| $ and save their phase information $<{\bf{\hat y}}^{(m)}_{{\mathcal G}_k}> $ for signal recovery. The enhanced spectrum amplitude $|{\bf{\hat s}}_{{\mathcal G}_k}|_{est} $ of graph speech is obtained by subtracting the estimated magnitude spectrum of noise $|{\bf{\hat n}}_{{\mathcal G}_k}|_{est} $from the noisy graph signal spectrum in graph frequency domain
\begin{equation}\label {eq15}
|{\bf{\hat s}}^{(m)}_{{\mathcal G}_k}|_{est}  = |{\bf{\hat y}}^{(m)}_{{\mathcal G}_k}| - |{\bf{\hat n}}_{{\mathcal G}_k}|_{est},
\end{equation}
where the estimation $|{\bf{\hat n}}_{{\mathcal G}_k}|_{est} $ is obtained by averaging the magnitude of the graph spectrum of $|{\bf{\hat n}}_{{\mathcal G}_k}|$ in the non-speech activity area. After adding $<{\bf{\hat y}}^{(m)}_{{\mathcal G}_k}>$ to the enhanced graph speech amplitude spectrum $|{\bf{\hat s}}^{(m)}_{{\mathcal G}_k}|_{est}  $, we convert it to $ {{\bf s}^{(m)}_{{\mathcal G}_{k} est}} $ in graph domain through (\ref{eq12}). Note that the function of our graph ${\cal G}_k$ is mainly to define the topology of the graph speech signal. Thus we can further obtain the corresponding frame ${s}^{(m)}_{est}$ in time domain by discarding the graph topology, and further obtain $s_{est}$.

\begin{algorithm}[t]
\setstretch{1.2}
\caption{Iterative graph spectral subtraction (IGSS)}
\label{algIGSS}
\begin{algorithmic}[1]
\REQUIRE noisy speech $y$; frame length $N$; step $k$ of the combined shift operator $\Psi_k$; noise threshold $\alpha$
\ENSURE the enhanced speech $s_{est}$
\STATE Divide $y$ into $M$ frames with frame length of $N$ and frame shift of $N/2$
\STATE Construct ${\mathcal G}_{k}=({\mathcal V},\Psi_k,\Psi_k)$
\STATE Estimate ${y}_{\alpha}$
\WHILE {${y}_{\alpha} \geq \alpha $}
\FOR{each frame $y^{(m)}$}
\STATE Construct the ${\bf y}^{(m)}_{{\cal G}_k}\text{{ indexed by }}{\mathcal G}_k$
\STATE Perform GFT on ${\bf y}^{(m)}_{{\cal G}_k}$ to obtain ${\bf{\hat y}}^{(m)}_{{\cal G}_k}$
\STATE  Save the phase $< {\bf{\hat y}}^{(m)}_{{\cal G}_k}> $, and extract the amplitude $| {\bf{\hat y}}^{(m)}_{{\cal G}_k}|^{'} $ of half vertices
\STATE Obtain the estimation of the noise $|{\bf{\hat n}}_{{\cal G}_k}|_{est}^{'}$ and subtract it to get $| {\bf{\hat s}}^{(m)}_{{\cal G}_k}|_{est}^{'}  = | {\bf{\hat y}}^{(m)}_{{\cal G}_k}|^{'} - |{\bf{\hat n}}_{{\cal G}_k}|_{est}^{'}$
\STATE Expand $| {\bf{\hat s}}^{(m)}_{{\cal G}_k}|_{est}^{'} $ to  $| {\bf{\hat s}}^{(m)}_{{\cal G}_k}|_{est} $
\STATE Recover $ {\bf{\hat s}}^{(m)}_{{\cal G}_k est}$ with the help of $< {\bf{\hat y}}^{(m)}_{{\cal G}_k}> $
\STATE Perform IGFT on  $ {\bf{\hat s}}^{(m)}_{{\cal G}_k est}$  to get  $ {\bf s}^{(m)}_{{\cal G}_k est}$ 
\STATE Obtain $ {s}^{(m)}_{est}$ by removing ${\mathcal G}_{k}$ of  $ {\bf s}^{(m)}_{{\cal G}_k est}$
\ENDFOR
\STATE Renew $y_{\alpha}$
\ENDWHILE
\STATE Restore $M$ frames to $s_{est}$
\RETURN ${s_{est}}$
\end{algorithmic}
\end{algorithm}

In particular, we observed that $\Psi_k$ used in this paper is a 0-1 real matrix in the form of row echelon. It is not a symmetric matrix where its eigenvalues are mainly complex and the real ones are very few, which can be ignored. Moreover, these complex eigenvalues are conjugated which means in the corresponding graph frequency domain, every amplitude appears twice. Since we just need to deal with the amplitude information, we only process the data of $N/2$ vertices and then utilize their conjugate properties to extend the $N/2$ amplitude coefficients to $N$ amplitude coefficients for greatly reducing the computation cost. The specific process of the GSS algorithm are shown in Algorithm \ref{algGSS}.

Based on GSS, we further propose the IGSS algorithm (see in  Algorithm \ref{algIGSS}) inspired by the idea of iteration\cite{i1,i2}. That is, to re-estimate the noise of the enhanced speech and perform the operation of graph spectrum subtraction until the estimated noise is less than a preset threshold $\alpha$, which is determined by averaging amplitude of the non-speech activity area in pure speech. If one is larger than $\alpha$, we consider that the speech is polluted and conduct GSS algorithm on the speech, otherwise, we consider that the signal too pure to be enhanced.

\section{Experimental Results}\label{S4}

We use  \textit{DARPA TIMIT Acoustic-Phonetic Continuous Speech Corpus (TIMIT)}\cite{b34}, which has a default sampling frequency of 16 $k$Hz as the dataset. We randomly selected 400 speech segments uttered by 40 speakers (20 male and 20 female), and each speaker has 10 speech segments. We use noise (white, pink and babble noise) from \textit{Standard noise NOISEX-92 library} \cite{b35} to generate our noisy signals with the signal to noise ratio (SNR) ranging from -15 to 15dB (step 5 dB). Each speech segment is framed by a rectangular window with a window length of 256 points and frame overlap of 50\%. Two traditional spectral subtraction methods, which are the basic spectral subtraction method (BSS) \cite{b23} and the iterative BSS method (IBSS), respectively, are used as benchmarks. The performance of each algorithm is measured via averaging the output SNR and Perceptual Evaluation of Speech Quality (PESQ)\cite{b31} of the corresponding 400 utterances.
\begin{figure}[htb]
\centerline{\includegraphics[width=0.95\linewidth]{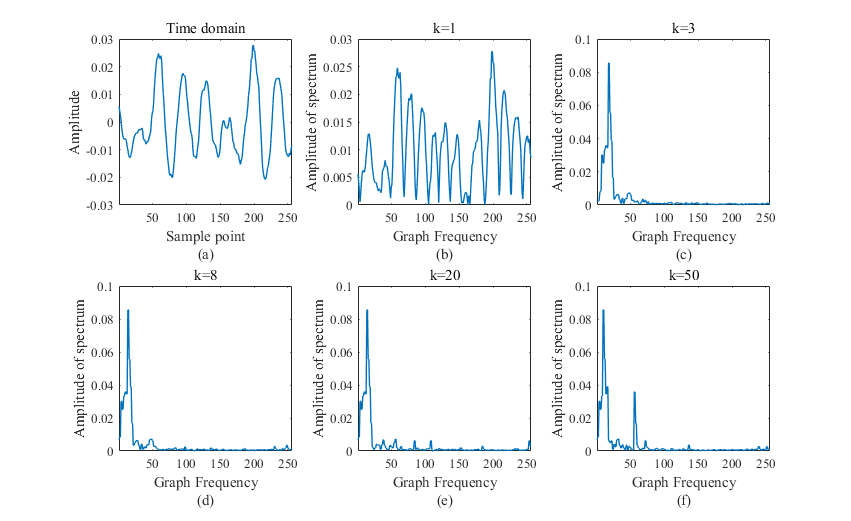}}
\caption{Spectrum of a pure speech frame with 256 points in different graph frequency domains. (a): an original speech frame in time domain; (b)-(f): spectrum of (a) in the graph frequency domain corresponding to $\mathcal G_1, \mathcal G_3, \mathcal G_8, \mathcal G_{20}, \mathcal G_{50}$, respectively. }
\label{figPureSpeechSpectrum}
\end{figure}

\begin{figure}[tbp]
\centerline{\includegraphics[width=0.95\linewidth]{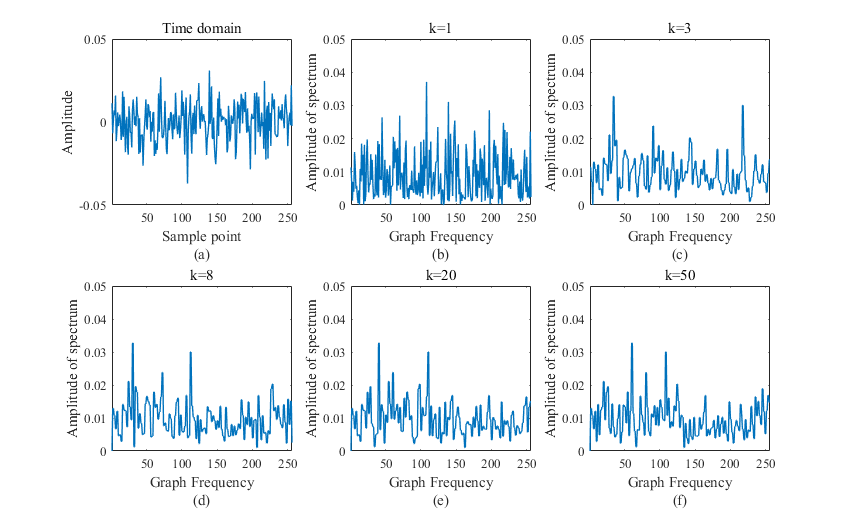}}
\caption{Spectrum of a white noise frame with 256 points in different graph frequency domains. (a): an original white noise frame in time domain; (b)-(f): spectrum of (a) in the graph frequency domain corresponding to $\mathcal G_1, \mathcal G_3, \mathcal G_8, \mathcal G_{20}, \mathcal G_{50}$, respectively. }
\label{figWhiteNoiseSpectrum}
\end{figure}

Fig.\ref{figPureSpeechSpectrum} shows the spectrum of a pure speech frame in different graph frequency domains corresponding to different $\Psi_k$, while Fig.\ref{figWhiteNoiseSpectrum} shows that of a white noise frame.
Note that in the case where $k=1$, $\mathcal G$ is equal to ${\mathcal G}_o$, which actually denotes the vertex domain. The corresponding graph Fourier matrix ${\mathcal F}_{\Psi_1}$ is then a unit matrix ${\bf{1}}^{N \times N}$. In other words, the graph frequency domain is equal to the graph domain. That is why the amplitude of spectrum of (b) in both Fig.\ref{figPureSpeechSpectrum} and Fig.\ref{figWhiteNoiseSpectrum} are the same as the absolute amplitude of (a). From Fig.\ref{figPureSpeechSpectrum} and Fig.\ref{figWhiteNoiseSpectrum}, we can further observe that in graph frequency domains corresponding to ${\mathcal G}_k$, the spectrum of the pure speech are mainly concentrated in low frequencies, while the distribution of white noise is relatively uniform, which is similar to distributions in the conventional Fourier domain. However, with the increase of $k$, we can obviously see the change of graph spectrum distribution for speech signal. The spectrum in middle frequencies show an increase, while the graph spectrum distribution of noise exists no obvious change. Considering that the difference between speech and white noise is more obvious for a small $k$, and a larger $k$ would lead to a higher computational cost, in this paper, we mainly discuss the case of $\Psi_3$.

\begin{figure}[t]
\centerline{\includegraphics[width=1\linewidth]{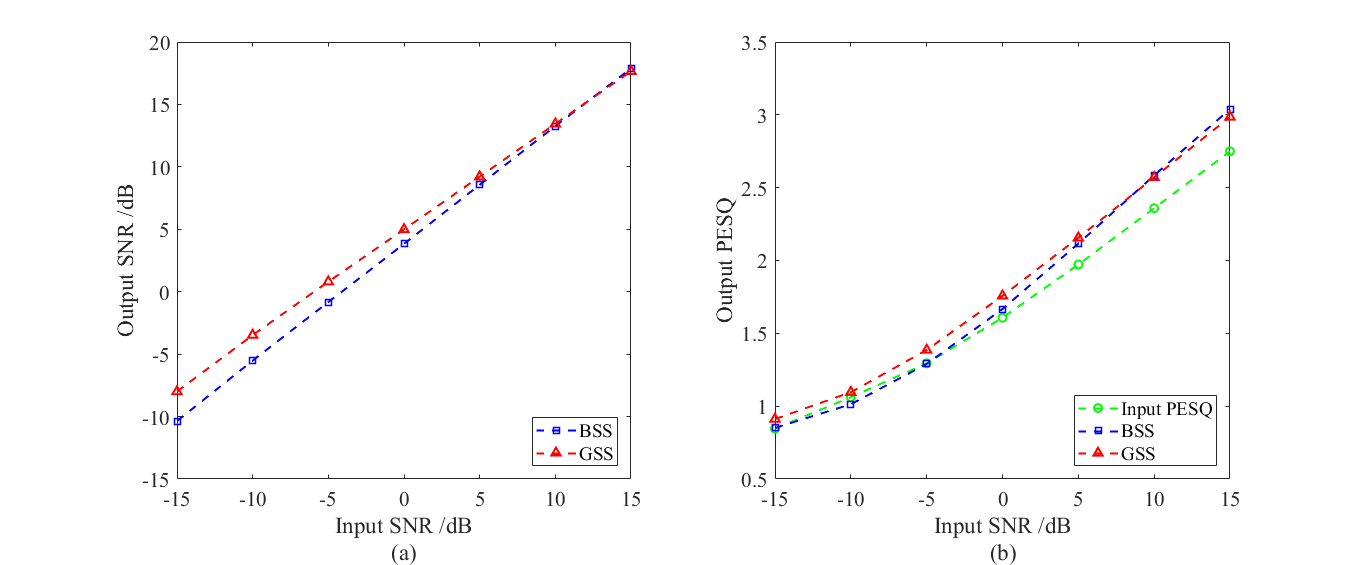}}
\caption{ The output SNR (a) and PESQ (b) of speech enhanced by proposed GSS method for white noise.}
\label{GSS}
\end{figure}

Fig.\ref{GSS} illustrates the SNR and PESQ of speech enhanced by BSS \cite{b23}and that of speech enhanced by GSS. In BSS\cite{b23}, there is a frame length of 256 points and a overlap of 50\%, and the Hamming windowed frames are analyzed with 256-point FFT. The noise is estimated from the first five frames of speech. From Fig.\ref{GSS} we can observe that  GSS outperforms BSS well in both SNR and PESQ for a low input SNR. Moreover, with the increase of input SNR, the improvement of GSS slightly slows down. This is because the quality of the input speech improves continuously under a high SNR, which means the representativeness of the noise in the non-speech activity area decreases correspondingly, thus GSS would remove too much details of speech.

\begin{figure}[t]
\centerline{\includegraphics[width=0.5\linewidth]{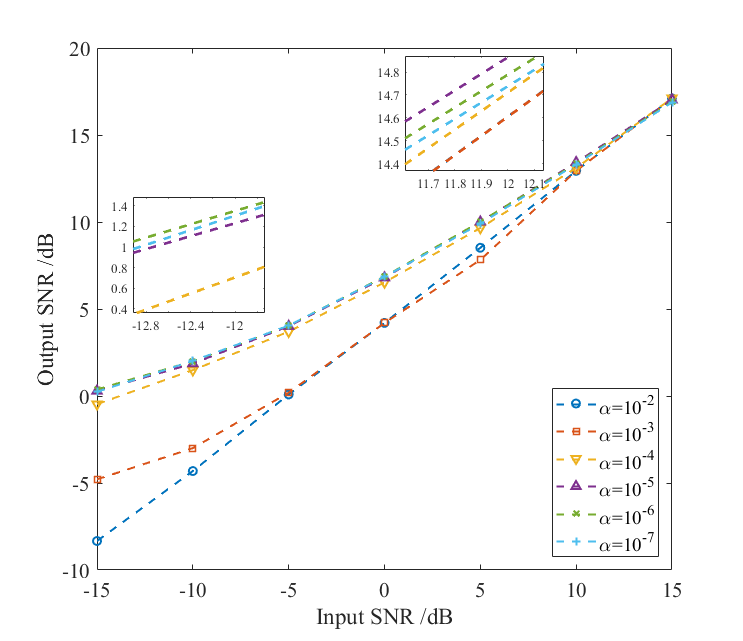}}
\caption{ The performance of IGSS on different $\alpha$ for white noise.}
\label{Alpha}
\end{figure}

To improve the performance of speech enhancement on white noise, we propose an IGSS algorithm and discuss the effect of iterative threshold $\alpha$. Fig.\ref{Alpha} shows the performance of IGSS on different $\alpha$. We can observe that the overall performance of IGSS increases significantly with the decrease of $\alpha$. When $\alpha$ goes down to $10^{-5}$, the enhanced performance starts to improve slowly, and the output SNR decreases slightly under a high input SNR. Although the performance of $\alpha=10^{-6}$ is slightly better than that of $\alpha=10^{-5}$ under low input SNR, it cannot compete with that of $\alpha=10^{-5}$ with the increase of input SNR. Note that the smaller $\alpha$ is, the more computation resources IGSS costs. Considering of both the overall enhanced performance and the computational cost, we set $10^{-5}$ to be our $\alpha$. 

\begin{figure}[ht]
\centerline{\includegraphics[width=0.9\linewidth]{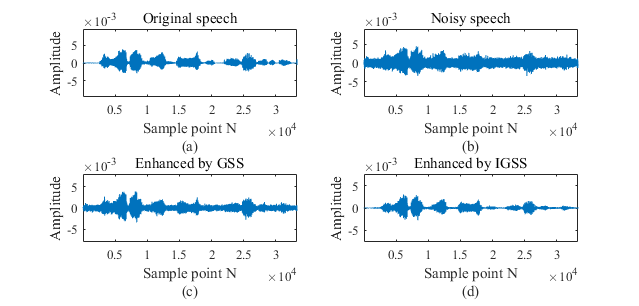}}
\caption{Time-domain waveforms of (a) original speech, (b) noisy speech, (c) enhanced speech by the proposed GSS algorithm, (d) enhanced speech by the proposed IGSS algorithm.}
\label{Waveform}
\end{figure}

Fig.\ref{Waveform} illustrates the time domain waveform of the original pure speech, noisy speech, speech enhanced by GSS and speech enhanced by IGSS, respectively. The noisy speech is interfered by white noise with an input SNR of 0 dB. From Fig.\ref{Waveform} we can see that both the GSS and IGSS significantly reduce the noise interference. Moreover, the proposed IGSS performs better than the proposed GSS in noise suppression, which can be attributed the idea of iteration in further suppressing the noise.

\begin{figure}[t]
\centerline{\includegraphics[width=1\linewidth]{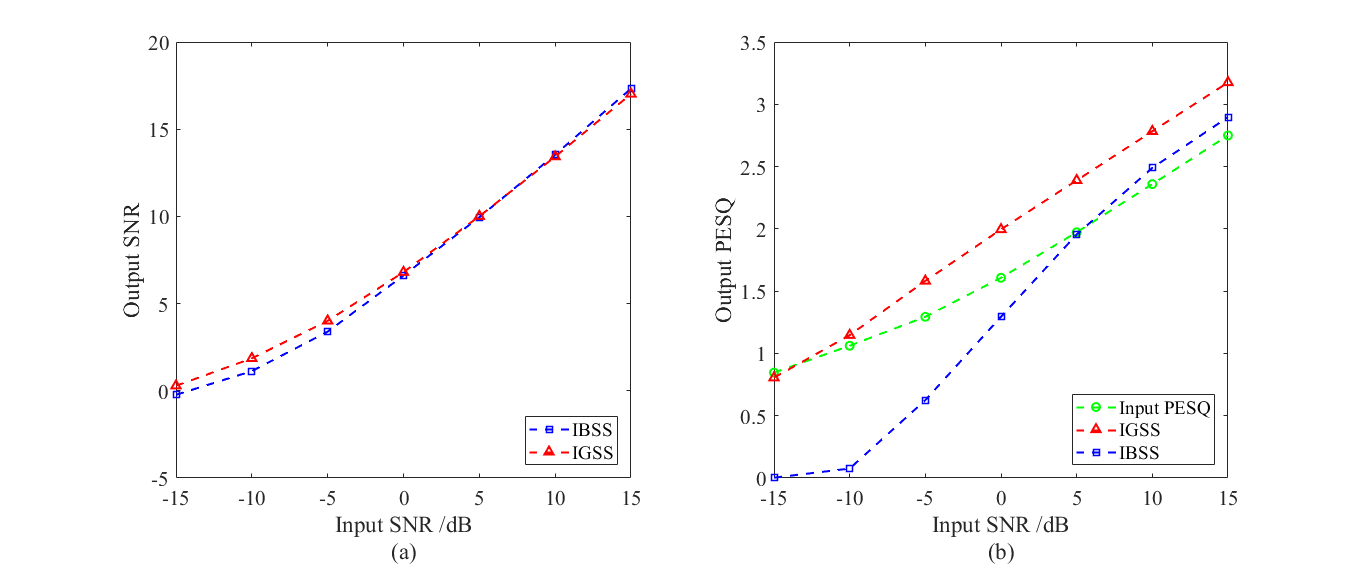}}
\caption{ The output SNR (a) and PESQ (b) of speech enhanced by IBSS and the proposed IGSS methods for white noise.}
\label{IGSSwhite}
\end{figure}

\begin{figure}[ht]
\centerline{\includegraphics[width=1\linewidth]{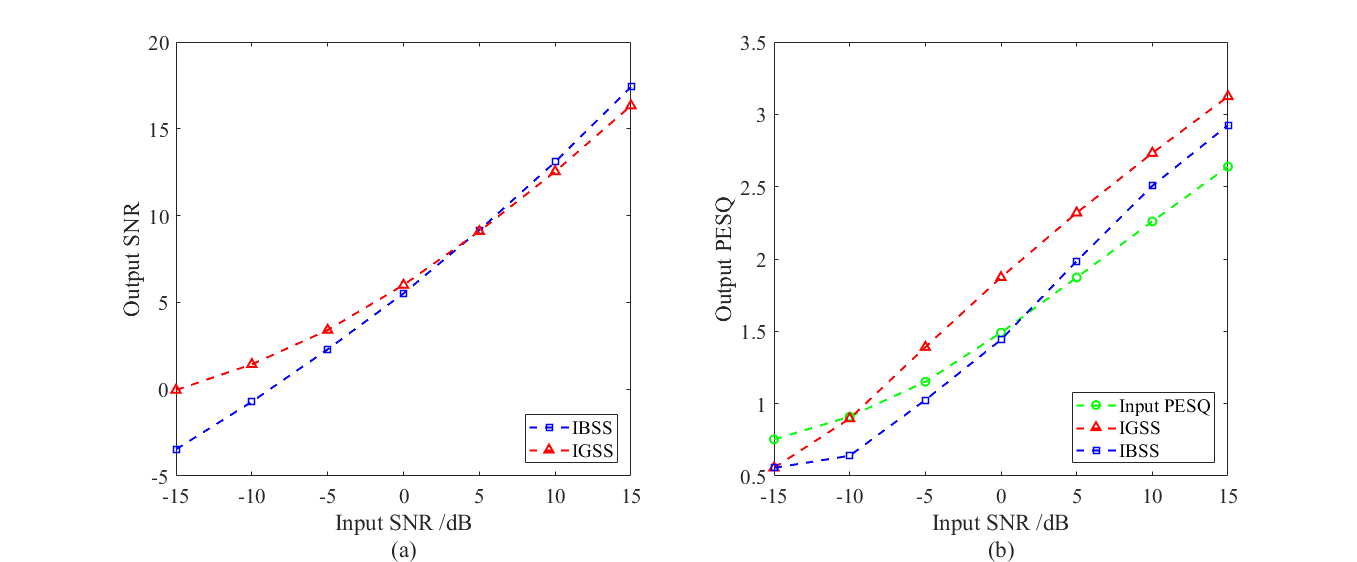}}
\caption{ The output SNR (a) and PESQ (b) of speech enhanced by IBSS and the proposed IGSS methods for pink noise.}
\label{IGSSpink}
\end{figure}

\begin{figure}[ht]
\centerline{\includegraphics[width=1\linewidth]{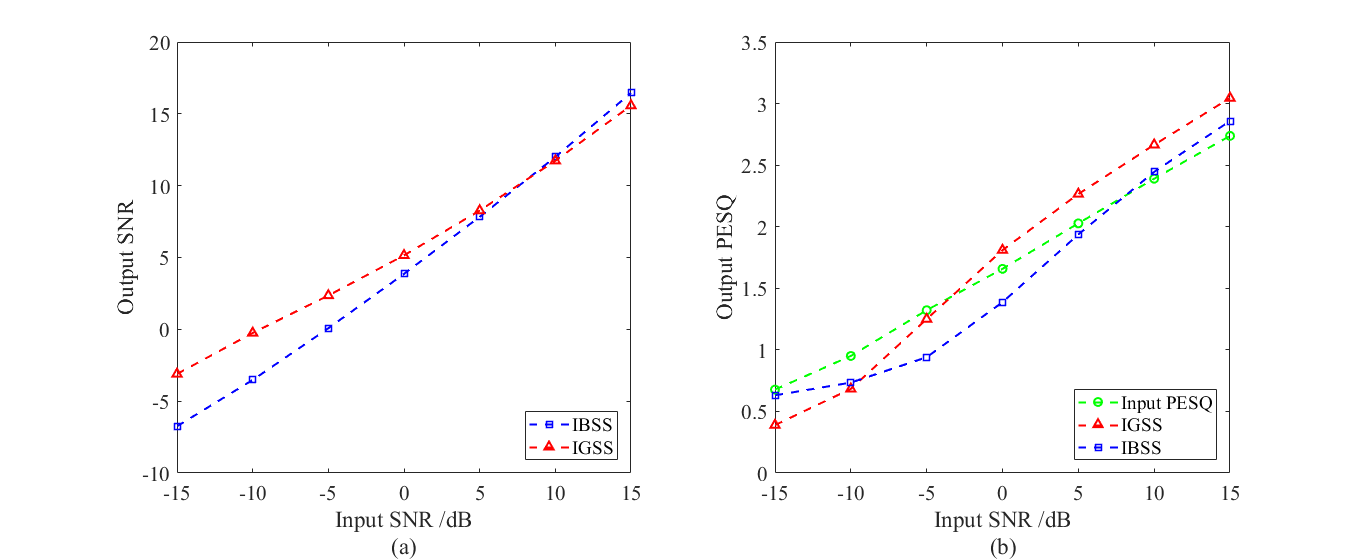}}
\caption{ The output SNR (a) and PESQ (b) of speech enhanced by IBSS and the proposed IGSS methods for babble noise.}
\label{IGSSbabble}
\end{figure}

In order to study the performance of IGSS algorithm on different types of noises, we carry out experiments on white, pink and babble noise.   Fig.\ref{IGSSwhite} ,Fig.\ref{IGSSpink} and  Fig.\ref{IGSSbabble} illustrate the performance of IBSS and IGSS algorithms in terms of the output SNR and PESQ on enhancing speech mixed with white, pink and babble noise, respectively. The iterative threshold $\alpha$ of both IGSS and IBSS is $10^{-5}$.

 Fig.\ref{IGSSwhite} shows the performance IBSS and IGSS algorithms on suppressing white noise. It can be seen from Fig.\ref{IGSSwhite}(a) that the proposed IGSS outperforms the IBSS in SNR especially at low SNR inputs. and the SNR improvement of IGSS is slightly lower than that of IBSS at high SNR inputs. In Fig.\ref{IGSSwhite}(b), we can observe that although the PESQ performance has a little bit decrease where SNR=15 dB, it shows significant improvements in PESQ for SNR $>$ -10 dB, while IBSS failed to improve the PESQ in conditions where SNR  $\leq$ 5 dB. This means that speech enhanced by IGSS is more in line with the auditory properties of the human ear.


 Fig.\ref{IGSSpink} shows the performance IBSS and IGSS algorithms on suppressing pink noise., We can observe from Fig.\ref{IGSSpink}(a) that the proposed IGSS has an obvious improvement than IBSS in pink noise suppression when the input SNR is lower than 5 dB. As the input SNR  continues to  increases, the improvements of IGSS slows down. However, we can observe that in Fig.\ref{IGSSpink}(b), the IGSS shows significant improvements in PESQ for SNR $>$ -10 dB, which illustrates that speech signals enhanced by IGSS method have better qualities. Moreover, IGSS failed to improve the PESQ just in poor SNR conditions where the input SNR $\leq$ -10 dB, while IBSS shows the improvement in PESQ only when  SNR is $\geq$ 5 dB. 

Fig.\ref{IGSSbabble} shows the performance IBSS and IGSS algorithms on suppressing babble noise. Note that babble noise used here is a non-stationary noise. We can observe from Fig.\ref{IGSSbabble}(a) that the proposed IGSS has an obvious improvement than IBSS in babble noise suppression when the input SNR is lower than 5 dB.  And the improvements in output SNR of IGSS begins to be slightly lower than that of IBSS for input SNR $\geq$ 10. In Fig.\ref{IGSSbabble}(b), we can observe that there exist no improvement for IBSS in PESQ when the input SNR is less than 10 dB, while IGSS begins to show the improvement in PESQ for SNR $\geq$ 0. In addition, IGSS outperforms IBSS in PESQ especially for high SNR inputs, which further illustrates that the quality of the enhanced speech signals by applying the IGSS method is better.

\section{Conclusions}\label{S5}
In this paper, we investigate the GSP based speech enhancement methods. We defined a set of combined $k$-shift operators to generate graphs and then using them to further construct graph speech signals.  By analyzing the distributions of speech and noise in different graph frequency domain, we propose the GSS to suppress the noise interference in noisy speech. Moreover, based on GSS, we propose IGSS to further improve the speech enhancement performance. Our experimental results show the superiority of GSS and IGSS algorithms in terms of both SNR and PESQ.\\

\bibliography{mybibfile}

\end{document}